\documentclass[sigconf, authorversion, nonacm]{acmart}

\AtBeginDocument{%
  \providecommand\BibTeX{{%
    \normalfont B\kern-0.5em{\scshape i\kern-0.25em b}\kern-0.8em\TeX}}}

\setcopyright{acmcopyright}
\copyrightyear{2023}
\acmYear{2023}
\acmDOI{XXXXXXX.XXXXXXX}

\acmConference[MM '23]{the 31st ACM International Conference on Multimedia}{October 28--November 3, 2023}{Ottawa, Canada}
\acmPrice{15.00}
\acmISBN{978-1-4503-XXXX-X/18/06}

\usepackage{listings}
\usepackage{xcolor}
\usepackage{subfig}
\usepackage{threeparttable}
\usepackage{diagbox}
\usepackage{comment}
\usepackage{listings}
\usepackage{xcolor}
\usepackage{nameref}
\usepackage{algorithm}
\usepackage{algpseudocode}
\usepackage{ragged2e}

\newcommand\JSONnumbervaluestyle{\color{blue}}
\newcommand\JSONstringvaluestyle{\color{red}}

\newcommand{\Sample}{\textbf{Sample}}
\newcommand{\Update}{\textbf{Update}}
\newcommand{\Init}{\textbf{Init}}

\newif\ifcolonfoundonthisline

\makeatletter

\lstdefinestyle{json}
{
  showstringspaces    = false,
  keywords            = {false,true},
  alsoletter          = 0123456789.,
  morestring          = [s]{"}{"},
  stringstyle         = \ifcolonfoundonthisline\JSONstringvaluestyle\fi,
  MoreSelectCharTable =%
    \lst@DefSaveDef{`:}\colon@json{\processColon@json},
  basicstyle          = \ttfamily,
  keywordstyle        = \ttfamily\bfseries,
  frame               = single,
}

\newcommand\processColon@json{%
  \colon@json%
  \ifnum\lst@mode=\lst@Pmode%
    \global\colonfoundonthislinetrue%
  \fi
}

\lst@AddToHook{Output}{%
  \ifcolonfoundonthisline%
    \ifnum\lst@mode=\lst@Pmode%
      \def\lst@thestyle{\JSONnumbervaluestyle}%
    \fi
  \fi
  \lsthk@DetectKeywords%
}

\lst@AddToHook{EOL}%
  {\global\colonfoundonthislinefalse}

\begin{document}

\title{LooPy: A Research-Friendly Mix Framework for Music Information Retrieval on Electronic Dance Music}

\author{Xinyu Li}
\affiliation{%
  \institution{New York University}
  \city{New York}
  \country{USA}
}
\email{xl3133@nyu.edu}

\renewcommand{\shortauthors}{Li, et al.}

\definecolor{light-gray}{gray}{0.95}
\newcommand{\code}[1]{\colorbox{light-gray}{\texttt{#1}}}
\begin{abstract}
  Music information retrieval (MIR) has gone through an explosive development with the advancement of deep learning in recent years. However, music genres like electronic dance music (EDM) has always been relatively less investigated compared to others. Considering its wide range of applications, we present a Python package for automated EDM audio generation as an infrastructure for MIR for EDM songs, to mitigate the difficulty of acquiring labelled data. It is a convenient tool that could be easily concatenated to the end of many symbolic music generation pipelines. Inside this package, we provide a framework to build professional-level templates that could render a well-produced track from specified melody and chords, or produce massive tracks given only a specific key by our probabilistic symbolic melody generator. Experiments show that our mixes could achieve the same quality of the original reference songs produced by world-famous artists, with respect to both subjective and objective criteria. Our code is accessible in this repository\footnote[1]{Code: \url{https://github.com/Gariscat/loopy}} and the official site of the project is also online\footnote[2]{Site: \url{https://loopy4edm.com}}.
\end{abstract}

\ccsdesc[500]{Information systems~Music retrieval}
\ccsdesc[500]{Applied computing~Sound and music computing}

\keywords{synthetic data generator, music information retrieval}


\begin{teaserfigure}
  \includegraphics[width=\textwidth]{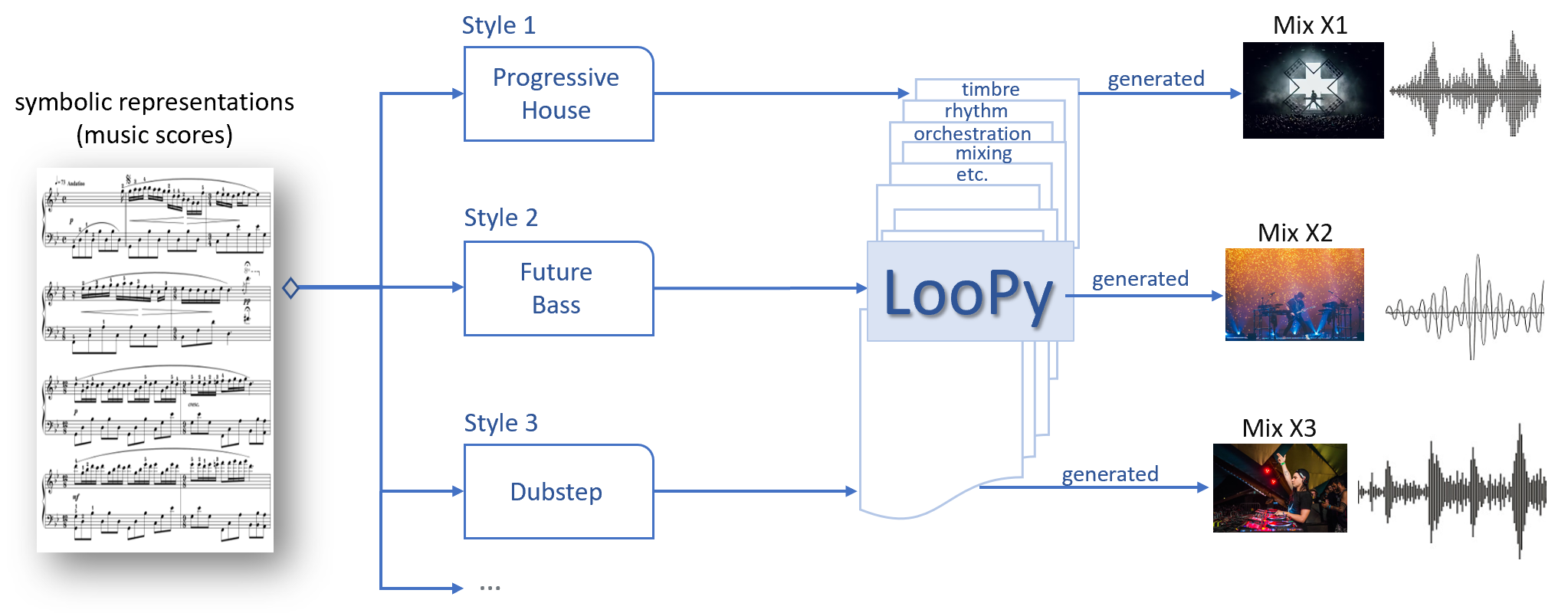}
  \caption{A conceptual diagram of our package "LooPy".}
  \Description{Concept}
  \label{fig:teaser}
\end{teaserfigure}

\maketitle

\section{Introduction}

AI-driven music generation has always been a popular research topic. Symbolic music generation has gone through and is still undergoing explosive development, with works like \cite{guo2022domainknowledgeinspired, han2022symbolic, Sulun_2022} addressing both some new methods and some of its applications. From naive sequential models like LSTM \cite{li_2021}, transformers \cite{score_transformer, yu2022museformer} to more sophisticated models, including Music VAE \cite{MusicVAE}, generative adversarial networks (GAN) \cite{MuseGAN} and diffusion models \cite{DiffMusic}, they are either trained in a supervised manner or via reinforcement learning \cite{RNN_RL}, which are able to generate symbolic music scores that achieve the level of human songwriters. However, music audio generation lags behind, since few works succeed to build scalable frameworks for music arrangement and mixing achieving the level of commercial production. Thus, it is relatively hard to automatically derive labels of music audio for downstream MIR tasks. We believe the development of symbolic music generation makes data synthesis from pre-determined arrangement-and-mixing templates a promising option. With the format properly aligned, researchers could easily synthesize a large amount of professional or quasi-professional EDM songs with such a template. In this manner, these songs are naturally annotated since the symbolic music representations are immediately available.

Our package aims to build a Python-based EDM data generator as a template framework, to convert generated symbolic music into well-produced audio files. Considering its popularity in radios and music festivals, we currently focus on the template of House music. Our provided demo templates could render professional-level EDM tracks considering both objective criteria (loudness, etc.) and subjective evaluations by human listeners. Note that this work is oriented for discriminative MIR tasks \underline{instead of} generative purposes, though it might also assist songwriting by human musicians. We would discuss the motivation in details in the following section. In short, our main contributions are three-fold as follows:
\begin{enumerate}
    \item We presented an open-sourced Python-based EDM generation framework with explicit musical score and mixer channel information, both as a data synthesizer for downstream MIR tasks (source separation and transcription, etc.), and also a potential tool for music generation and its applications.
    \item We built a ready-to-use template for professional-level EDM generation using this framework as an example, achieving the level of commercial production by experts with both objective and subjective criteria.
    \item We provided a trivial (rule-based) rhythm and melody generator that produces massive symbolic music scores given some key and chord progression, validating our hypothesis of large-scale data synthesis.
\end{enumerate}

\section{Motivation}

In the field of music information retrieval (MIR), acquiring labelled data for technically complex music genres like electronic dance music (EDM) has always been intractable. To begin with, data annotation could hardly be carried out for EDM songs. Unlike traditional music with the piano and some orchestral instruments, or rock music with the guitar, the bass and the drums, professional EDM songs use several orders of magnitude more instruments and synthesizers. Also, EDM tracks use many automated effects like filter automations and reverberation automations\footnotemark[5], etc., making the mixer channels much more technically sophisticated than those of other music genres. Therefore, it is significantly more difficult to annotate EDM songs. Moreover, it is also unfeasible to acquire data from projects that render EDM songs. The EDM producer community use digital audio workstations (DAWs) like \cite{fl, ableton, logic}, and each track is characterized by a DAW project, instead of a raw music score. However, there are almost no DAW that has features mature enough for users to draw music notes onto the piano roll with scripts conveniently. One exception is Reaper with its built-in scripting feature named ReaScript \cite{ReaScript}. Still, considering its relative unpopularity in the EDM community, it might be challenging for researchers and musicians to collaborate on templates using Reaper. Therefore, even with numerous symbolic music generated by advanced deep learning algorithms, we could hardly mass-produce EDM tracks given some DAW project template, not to mention the difficulty of building a high-quality template itself. In all, datasets or data generator with detailed musicological annotations (like note positions, durations, etc.) for EDM songs remain to be a vacancy.
\footnotetext[5]{Brooks masterclass: \url{https://www.youtube.com/watch?v=JTJwR125DGQ}}

In the face of all these difficulties, why bother to retrieve detailed information for electronic dance music? We propose several application scenarios to show the significance of MIR for EDM songs. Firstly, with more fine-grained information (timbre, rhythm, etc.) retrieved, music streaming services could build more accurate recommendation system, providing better tailored recommendations for their users. Secondly, it can enhance the performance of music visualization systems. Previous work like \cite{housex} has demonstrated the potential of visualization automation with supervised learning methods, given only relatively coarse retrieved information (sub-genres). Its full power can be unleashed only if the system can retrieve more fine-grained information. Thirdly, automatic music transcription (AMT) is uninvestigated for EDM songs, with current state-of-the-art transcription algorithm \cite{gardner2022mt} trained on datasets like Slakh2100 and \cite{Slakh2100} and MusicNet \cite{MusicNet} that only contain classical music or other genres played by the piano, the guitar, the bass and the drums. Last but not least, music source separation (MSS) is not mature for EDM songs, but plays a huge part in EDM festivals, enabling artists to mix different parts of different tracks in the show. Current MSS algorithms like \cite{rouard2022hybrid} are trained on datasets like Slakh2100 and MUSDB18 \cite{MUSDB18}, using the drum-bass-vocal-other scheme. Theses algorithms work to some extent for EDM, but are unable to separate detailed sources like leads and chords among instruments. With these application prospects, we present "LooPy" with its internal structure similar to FL Studio (one of the most popular DAWs in the community) and establish a bridge from symbolic music to professionally produced EDM, as an EDM data infrastructure for MIR research.

\section{Related Work}

Though relatively unexplored, MIR for EDM has been discussed by a few previous works. Hsu et al. \cite{hsu2021deep} proposed novel neural architectures for EDM sub-genres classifications. Zehren et al. \cite{zehren2020automatic} presented a method for "cue point" detection to automatically construct transitions among tracks as professional DJs do onstage. These works though demonstrated success on their respective sub-tasks, stopped with relatively coarse information retrieval. Other harder tasks like music transcription and music source separation remain barely addressed.

Now, we discuss several works that bear resemblance to ours. There are several other Python codebases or packages designed for music generation, that output audio waveforms instead of symbolic scores or MIDI files. Previous developers built a music programming framework "Musicpy" \cite{Musicpy} that features handy syntax and readability. Besides, a more recent project "DawDreamer" \cite{DawDreamer} blazed a trail for DAWs with Python, which supports core DAW features including virtual studio technology (VST) instruments and effects. Another Python package is “pedalboard” by Spotify \cite{pedalboard}, featuring fast I/O utilities and adding sound effects, along with compatibility with TensorFlow \cite{tensorflow2015-whitepaper} data pipeline. Though there works do not cover our idea, they do show a promising prospect of music audio manipulation using Python scripts. Our codebase also uses several effects of theirs as high-level wrapped methods for sound mixing.

Our work "LooPy" features audio generation, given instructions and symbolic music representations. In other words, it could be viewed as a DAW to some extent as well. However, our work aims to serve as a high-quality template framework for EDM songs. For simplicity, generators (instruments) in "LooPy" directly use sounds pre-rendered in DAWs like FL Studio 20 \cite{fl} with VST plugins like Sylenth 1 \cite{sylenth1}, rather than synthesize waveforms from scratch. Our priority is to provide credible EDM track templates with a Python API that are compatible with symbolic music scores. If all goes well, we plan to incorporate more generic DAW functions, like loading VST plugins, to offer more choices for users.

\section{Features}

\begin{figure}[!htp]
    \includegraphics[width=8.5cm]{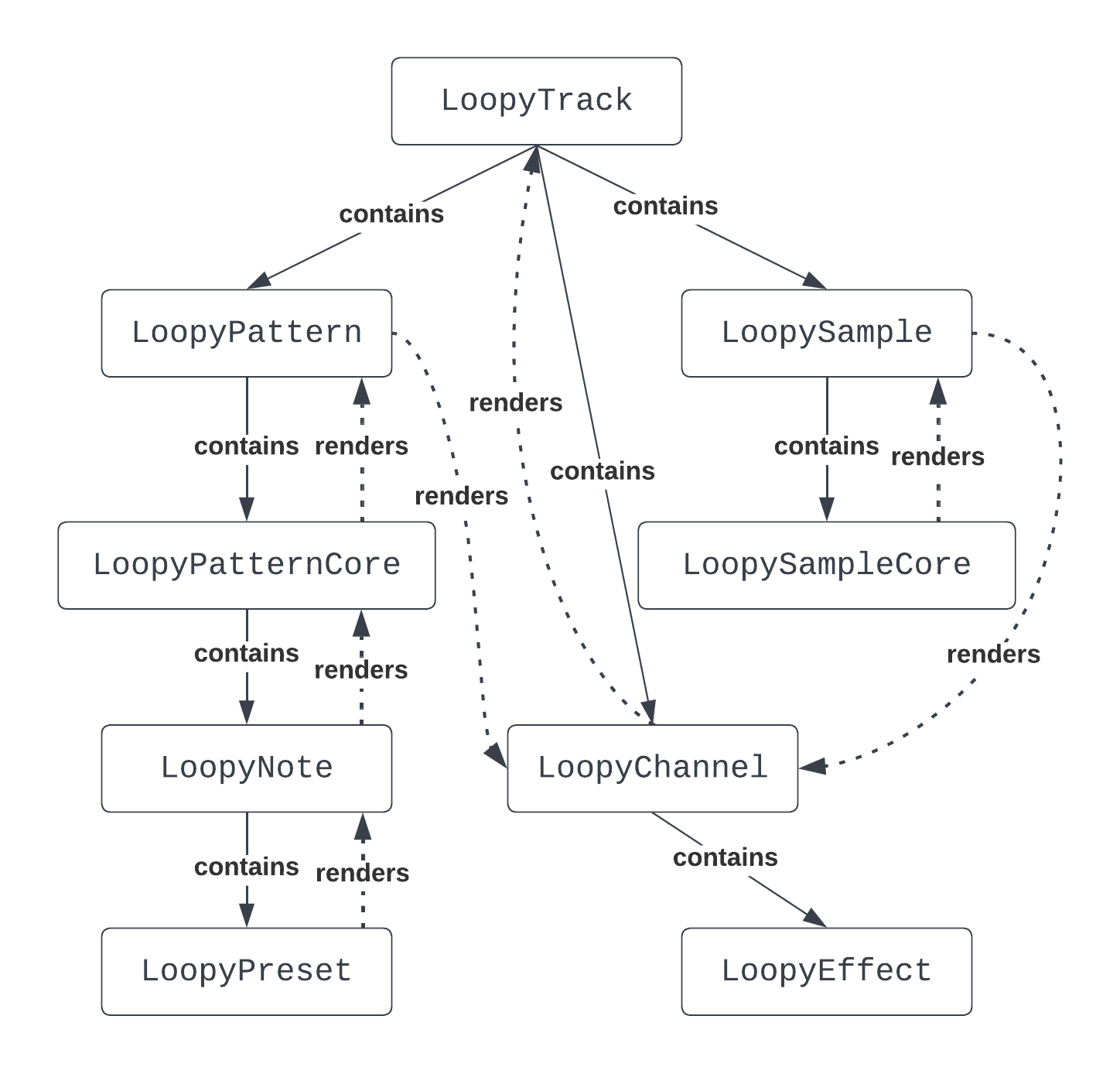}
    \caption{The core classes of the package and their relations.}
    \label{fig:scheme}
\end{figure}

Wrapped up as a Python package, "LooPy" features several modules that are essential to a quick, convenient and professional construction of an EDM track template. In the following sub-sections, we introduce the functions supported by this package.

\subsection{Core Classes}
The data of a track is stored in dictionaries hierarchically in "LooPy", which is built upon several core classes. They are shown as the rectangles in Fig.\ref{fig:scheme}. The next few paragraphs demonstrates how all these classes work together as a whole equivalent to a complete EDM project file. In Appendix \ref{appendix:setup}, we demonstrate detailed connections between our work and FL Studio \cite{fl}, which would help users understand our setup better if they are familiar with FL Studio or any other DAW designed for EDM production.

Each track (song) is represented as a \code{LoopyTrack} instance. Since EDM songs feature loops of certain musical structures (melody, chord progression, etc.), each type of repeating musical paragraph corresponds to a \code{LoopyPatternCore} instance. For each occurrence of this paragraph, a \code{LoopyPattern} instance is stored by the track. In other words, one pattern core defines a type of pattern and a pattern defines an actual paragraph with its position in the track (measure index). Similarly, a track stores all the occurrences of samples and each occurrence contains its sample core that defines the path storing the audio sample. Besides, a track also stores all the mixer channels used in the rendering process.

Inside the pattern core, all the notes are stored. Each note stores the pointer to the generator that generates the note. In "LooPy", all the generators are currently defined as \code{LoopyPreset} objects instead of waveform synthesizers that generate sounds from scratch. We pre-rendered the sounds of 88 pitches on a piano keyboard, for each preset. Therefore, when a note is defined, we directly load the audio from the pre-rendered file. Of course, we would consider integrating some waveform synthesizers to provide more freedom of sound design.

To render the audio of a track, the classes work in a manner similar to back propagation. The audio of each note is rendered by the preset. The audio of a pattern core is rendered by the assemblage of the rendered notes. The audio of a pattern/sample is basically the audio of a pattern/sample core. Then, sounds of patterns and samples go into mixer channels where each is defined as a \code{LoopyChannel} instance. Each channel contains several effects like EQ, compression, reverberation, etc. These effects are wrapped as \code{LoopyEffect} classes based on \cite{pedalboard}. Finally, the outputs of channels are assembled as the audio of a track.

\subsection{Styles}

Each artist, or sometimes each released track, exhibits some unique styles in composition and mixing. We built a \code{LoopyStyleBase} class as the base class for EDM styles or sub-genres. It contains two dictionaries, a sound style sheet and a channel style sheet, including all the information needed to compose and mix the track. Then, each different style is extended from the \code{LoopyStyleBase} by modifying the two class variables, the sound sheet and the channel sheet.

Examples of the sound sheet and the channel sheet are given in Listing \ref{sound_sheet} and \ref{channel_sheet}. For simplicity in saving and loading (as JSON files), we format both sheets as Python dictionaries. For each key-value pair in the sound sheet, the key is a part (e.g. bass, kick) of the track while the value is a list of dictionaries, where each contains the path to the source file and the parameters of loading the sound. Note that for effect samples (fx) like fills and downlifters, we put a dozen of samples in the directory and randomly choose a few to compose each track, to introduce more diversity and variety. The channel sheet, on the other hand, simply maps a part of the track to a list of dictionaries, where each stores the name and parameters of the \code{LoopyEffect} objects.

\begin{lstlisting}[style=json, label=sound_sheet, caption=the sound sheet]
"bass": [
  {"source_path": "Ultrasonic-BS-Home.wav",
   "gain": -19.0},
   ......
],
"kick": [
  {"source_path": "KSHMR_Top_Kick_03.wav",
   "gain": -21.1,
   "blank_every": 8},
   ......
],
"fx": [
  {"type": "main-fill",
   "dir": "../samples/main-fill",
   "highpass": 250,
   "gain": -12,
   "every": 8},
   ......
],
......
\end{lstlisting}

\begin{lstlisting}[style=json, label=channel_sheet, caption=the channel sheet]
"lead": [
  {"type": "highpass",
   "freq": 300},
  {"type": "sidechain",
   "attain": 0.5,
   "interp_order": 3,
   "mag": 0.66},
  {"type": "reverb",
   "dry_level": 0.5,
   "wet_level": 0.8},
  {"type": "balance",
   "gain": 1.5},
  {"type": "limiter",
   "thres": -6.0}
],
......
\end{lstlisting}

\subsection{Generation}
To generate the audio of the track, "LooPy" constructs the presets and loads the samples into the track according to the sound sheet, then constructs the mixer channels according to the channel sheet. With the input of music scores (the information of notes), a bunch of \code{LoopyNote} instances are initiated and placed into \code{LoopyPatternCore} objects. Each \code{LoopyNote} object contains the address of the \code{LoopyPreset} object that "plays" the note, and other information of the note such as position and note value, etc. After all patterns and samples are properly placed, the track is ready for rendering.

Now that the mixing tasks have been solved, how do we derive massive symbolic music representations? Symbolic generation has been a deeply studied field and is still undergoing intensive research. Since time is limited, we implemented a deterministic rhythm melody generator given some chord progression, alongside the main modules mentioned in previous sub-sections.

In practice of melody writing, the pitch and duration of a note should not be disentangled. However, since our work is primarily oriented as an infrastructure for discriminative tasks, we separate these two features to simply the symbolic generation process, at the expense of composition quality. We first generate a rhythm following the steps in Alg. \ref{alg:rhythm}, getting a list of "placeholders" of notes. Each placeholder indicates the position and duration of a note. Then, we randomly select pitches in the specified major/minor scales and put them into the placeholders, to get a list of notes. Each item in this list is $(k,\ v,\ st)$ where $k$ is the pitch, $v$ is the duration (value) and $st$ is the position.

\begin{figure}[!htbp]
    \includegraphics[width=\linewidth]{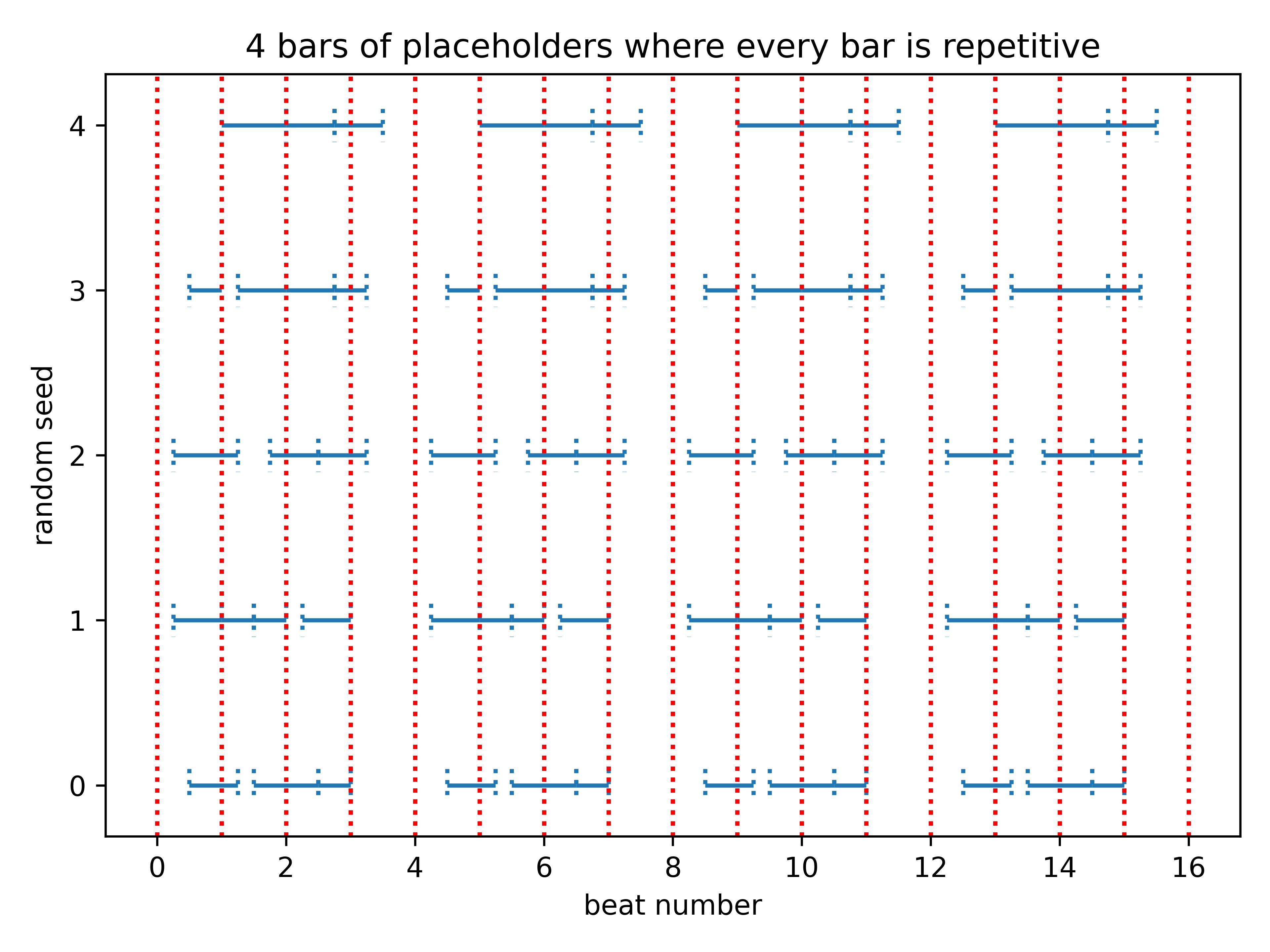}
    \bigskip 
    \includegraphics[width=\linewidth]{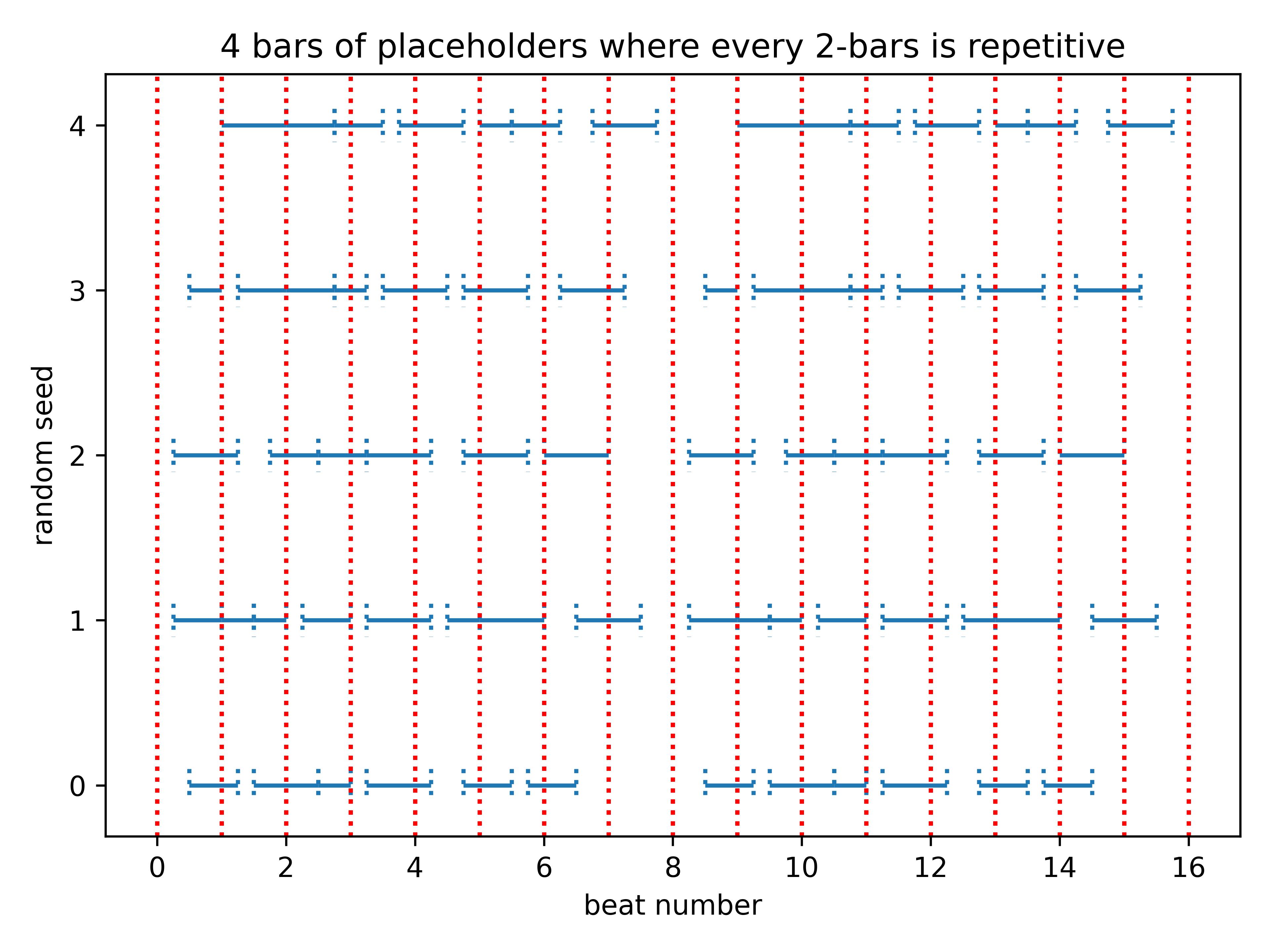}
\caption{Examples of rhythm generation results}
\label{fig:rhythm}
\end{figure}

Some generated rhythms are show in Fig. \ref{fig:rhythm}. Each blue segment marks a placeholder to be filled by a note and each red vertical dotted line marks a beat. We generate rhythms for 4 bars, since most melodies would sound recurrent every $n$ bars where $n$ is no bigger than 4. Furthermore, we provide a parameter $rep\_bar$, which is the smallest number of bars such that the melody rhythm is repetitive, for users to imitate the styles of their reference tracks.

\begin{algorithm}
\caption{A rule-based algorithm to generate rhythm for $N$ bars with time signature 4/4}
\label{alg:rhythm}
\begin{algorithmic}
\Require $vs,\ r,\ \lambda$ \Comment{list of note values such as [$\frac{1}{16}, \frac{1}{8}, \frac{3}{16}, \frac{1}{4}$], resolution (value of the shortest note) such as $\frac{1}{16}$, parameter of Poisson distribution $\text{Pois}(\lambda)$}
\State \Init $\ st, ed \gets -1, 0$ \Comment{start and end positions of a placeholder}
\State \Init $\ ps\gets []$
\While{true}
    \State \Sample $\ t\sim\text{Pois}(\lambda)$
    \State \Update $\ st \gets ed + t\cdot r$
    \State \Sample $\ v\sim\text{Unim}(vs)$ \Comment{choose a note value (uniformly as default)}
    \State \Update $\ ed \gets st + v$
    \If{$ed>N$}
        \State \textbf{break}
    \EndIf
    \State \Update $\ ps \gets ps + [(v,\ st,\ ed)]$ \Comment{append this placeholder}
\EndWhile
\Ensure $ps$ \Comment{list of placeholders}
\end{algorithmic}
\end{algorithm}

\begin{algorithm}
\caption{A rule-based algorithm to generate notes for $N$ bars with time signature 4/4 from some rhythm placeholders}
\label{alg:melody}
\begin{algorithmic}
\Require $ps,\ ks,\ ws$ \Comment{list of placeholders, list of note pitches such as [C5, D5, E5, G5, A5, B5, C6], probability weights (default as uniform)}
\State \Init $\ ns \gets []$
\For {$v,\ st,\ ed\in ps$}
    \State \Sample $\ k\gets np.random.choice(ks,\ p=ws)$
    \State \Update $\ ns \gets ns + [(k, v, st)]$
\EndFor
\Ensure $ns$ \Comment{list of notes}
\end{algorithmic}
\end{algorithm}

\section{Experiments}

Our experiments consist of two parts. The first part focuses on remaking several popular EDM songs and the comparison between our remakes and the original tracks. The second part shows the evaluation results of our experimental massive parallel generation of certain EDM style.

\subsection{Remakes}

\begin{table*}[!htp]
  \begin{threeparttable}
  \caption{Loudness comparisons}
  \label{tab:loudness}
  \begin{tabular}{|c|c|c|c|c|c|}
    \hline
    \diagbox{Criteria}{Classes} & Style 1 Ours & Style 1 Original & Style 2 Ours & Style 2 Original & General Reference\\
    \hline
    \textit{LUFS integrated (dB)}& $\textbf{-5.4}$ & $\textbf{-5.3}$ & $\textbf{-4.9}$ & $\textbf{-5.6}$ & $\textbf{-8.0}$\\ 
    \hline
    \textit{LUFS short-term (dB)}& $-5.3\pm 0.2$ & $-5.3\pm 0.1$ & $-5.0\pm 0.2$ & $-5.6\pm 0.2$ &\diagbox{}{}\\
    \hline
  \end{tabular}
  \begin{tablenotes}
    \item[a] Style 1 refers to melodic progressive house, featuring Tobu\footnotemark[3]. Style 2 refers to emotional progressive house, featuring other two world-renowned producers\footnotemark[4].
    \item[b] "Ours" refers to our generated data and "original" refers to the original track which we intend to remake. All the readings are measured through an $8$-bar drop section (the climax of EDM tracks).
  \end{tablenotes}
  \end{threeparttable}
\end{table*}

\footnotetext[3]{Tobu - Life: \url{https://soundcloud.com/7obu/tobu-life}}
\footnotetext[4]{Starlight (Keep Me Afloat): \url{https://soundcloud.com/martingarrix/martin-garrix-dubvision-and-shaun-farrugia-starlight-keep-me-afloat}}

\begin{figure}[!htp]
    \includegraphics[width=8.5cm]{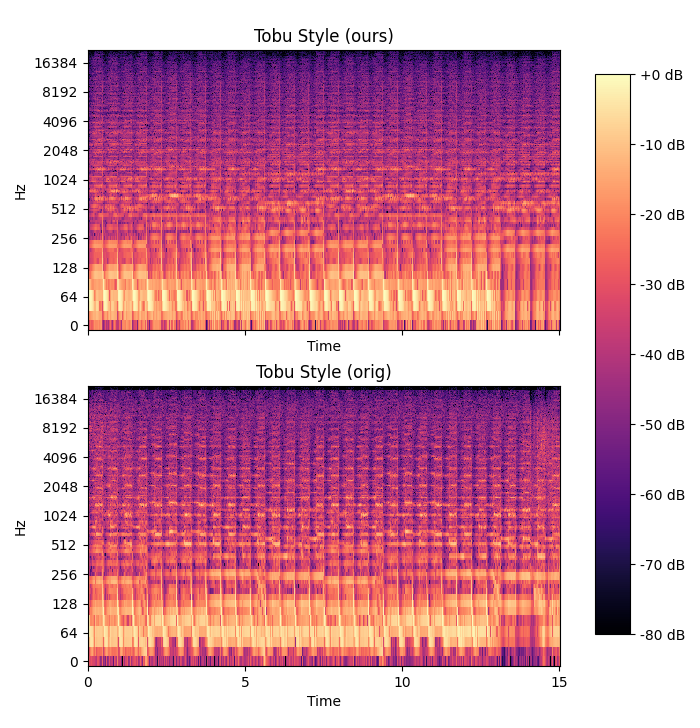}
    \caption{Mel-spectrograms of ours and the original track}
    \label{fig:spec_comp}
\end{figure}

In this subsection, we remaked the drop of two popular EDM tracks\footnotemark[3]\footnotemark[4]. The evaluation of the quality of an EDM track lacks a uniform quantitative standard. In terms of production, one common criterion is the loudness. Currently, the mainstream EDM songs aim for an integrated LUFS loudness no less than $-8.0$ dB \cite{mastering}. Moreover, many mainstream music nowadays approach to $-5.0$ LUFS at the loudest part and sometimes even higher, including two of our reference songs\footnotemark[3]\footnotemark[4]. Experiments show that our generation results could achieve strongly competitive loudness without perceptible distortion, compared to our reference songs, as shown in Tab. \ref{tab:loudness}. Also, from Fig. \ref{fig:spec_comp}, we could see that the melody and bassline of the reference track is slightly more outstanding from the perspective of mel-spectrograms. However, from the auditory perspective, our generated track also features a bright melody and a fat bassline. In fact, from the mel-spectrograms, our track has a fuller mix and that conform to the result that ours is a bit louder.

Admittedly, there are other criteria of production quality such as stereo width, etc. However, our work is positioned as a synthetic data generator for information retrieval, instead of a digital audio workstation. Whether the mix are loud and clean is an essential indicator of the production quality of EDM music. Adding other objective criteria would be considered as a primary future work.

\subsection{Batch Production}

We also carried out a massive generation experiments\footnotemark[6] using our primary template for progressive house. Following that, we conducted a double-blinded survey on examples of our generation results and some commercially released tracks, with respect to 3 criteria (creativity, naturalness and musicality). The evaluated six pieces are \href{https://drive.google.com/drive/folders/1ESC-KwPbElGKlfApz7AhGk3rgxWCYtDM?usp=sharing}{\color{blue}{here}} where the 1st, 3rd, 5th are generation examples and the 2nd, 4th, 6th are excerpts of commercial releases (the participants do not know this order). Without introducing distortions, we randomly pitched up or down the commercial excerpts to avoid the potential use of music recognition software.

\begin{figure}[!htp]
    \includegraphics[width=8.5cm]{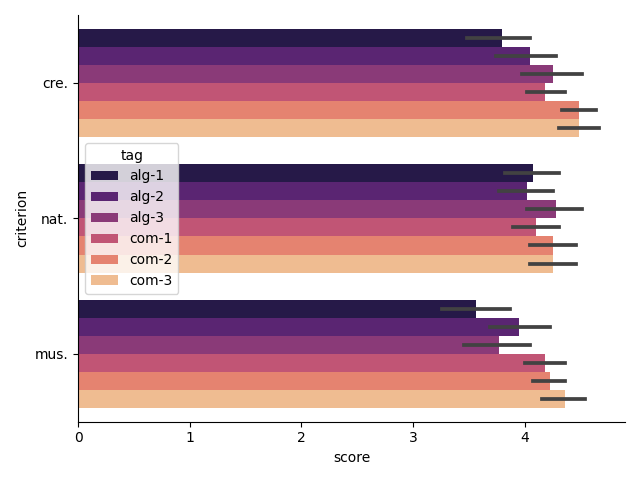}
    \caption{Subjective evaluation results}
    \label{fig:subjective}
\end{figure}

The evaluation results are given in Fig. \ref{fig:subjective}, which shows that the musicality and creativity of commercial releases are still significantly better than pieces generated by our trivial algorithm. However, the mixing template makes the naturalness of our tracks close to that of commercial releases. Considering the position of "LooPy" as an infrastructure for discriminative MIR tasks, the naturalness is the most important criterion here since it is strongly correlated to the production quality of the track, while creativity and musicality are mainly associated with composition. Our deficiency of thses two criteria should be mitigated by using state-of-the-art symbolic music generation algorithms like \cite{han2022symbolic, yu2022museformer}.

\footnotetext[6]{All renders and data are in this folder: \url{https://drive.google.com/drive/folders/1X-jArl_6DsBxZdXGL7wzgaVI4m6f8wiy?usp=sharing}}

\section{Conclusion}

To sum up, we propose a research-friendly data framework, an audio generator for music information retrieval (MIR) focusing on electronic dance music (EDM). Users could generate a piece of well produced music audio from specified symbolic melody line and chord progression or produce a large amount of tracks specifying a key, with less than $10$ lines of code. Experiments show that the generator could output professional-level music compared to mainstream EDM tracks. The following paragraphs present several aspects where future efforts should be made.

Firstly, the criteria used in our assessment are not perfectly inclusive. Though our current generation results sound reasonably good, we cannot ensure they meet all the industry standards by now. To make our generated data more reliable for downstream MIR tasks, other quantitative criteria such as stereo width should be considered to guarantee that models trained using our data could generalize well to actual commercial music in the real-world industry.

Secondly, more genres should be covered by our template. There is no doubt that other genres like future bass, electro house, dubstep, etc., also occupy a huge portion of the market. Even for progressive house, our template only covers one of its popular styles. The construction of templates for these genres and styles is an essential step for this package to be widely used in the community. Admittedly, it is unrealistic for us to build a template for every popular EDM genre all by ourselves. That is also the reason why we spared no efforts to make this package high-level so professional artists and MIR researchers can collaborate easily with it. We plan to invite others professionals for the construction of new templates as one of our next steps.

Last but not least, some functions in "LooPy" need to be improved and extended. There are several functions that we found unwieldy during our experiments. For example, routing the output of a channel to the input of another channel is inconvenient by our currently codebase. Such routing function would make the mixer channels much more concise, since users could create one sidechain channel for multiple instrument channels rather than add a sidechain effect on each channel. Besides, advanced features such as automations (for volume, filters and reverberation, etc.) need to be built to make the tracks more professional, introducing more dynamics. We also plan to make third-party VST plugins compatible with our effect modules in future. There advanced functions would make "LooPy" a more standard and generalizable infrastructure for MIR.

\begin{acks}
We thank the anonymous makers of the Sylenth1 sound banks that we used to render our presets, without which we could not easily build the sound bank of our generators. We also thank the authors of "pedalboard" for open-sourcing their code, making the construction of our effect modules much more straightforward.
\end{acks}

\bibliographystyle{ACM-Reference-Format}
\bibliography{ref}

\appendix

\section{Setup}
\label{appendix:setup}

\begin{figure}[!htbp]
    \includegraphics[width=8cm]{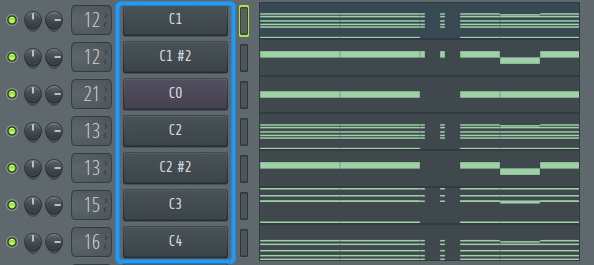}
    \caption{The list of generators}
    \label{fig:generators}
\end{figure}

\begin{figure}[!htbp]
    \includegraphics[width=8cm]{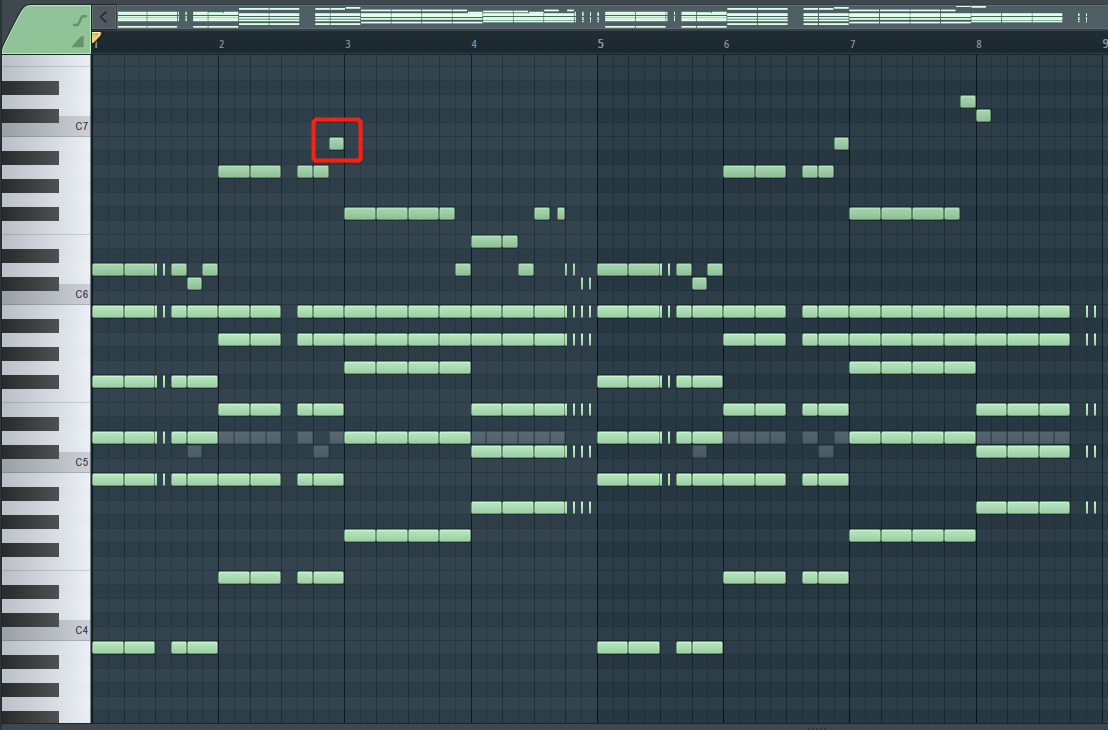}
    \caption{The piano roll of a pattern}
    \label{fig:pattern_core_notes}
\end{figure}

\begin{figure}[!htbp]
    \includegraphics[width=8cm]{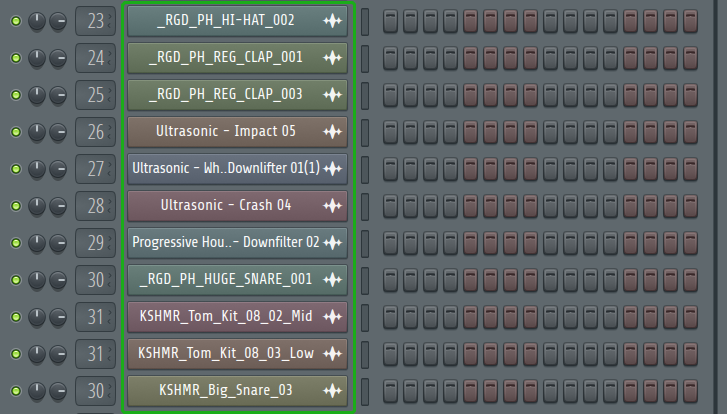}
    \caption{The list of samples used in the track}
    \label{fig:sample-cores}
\end{figure}

\begin{figure}[!htbp]
    \includegraphics[width=8cm]{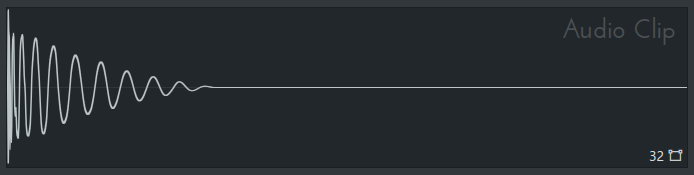}
    \caption{The waveform of a sample}
    \label{fig:sample-core}
\end{figure}

Our work is strongly inspired by the architecture of FL Studio \cite{fl}, thus the setup of "LooPy" is similar to that of FL Studio. This appendix section shows how the major functionality of FL Studio is implemented in our package.

As shown in Fig. \ref{fig:generators}, instruments are placed as generators in FL Studio. In "LooPy", for simplicity, we rendered some common EDM sounds from Sylenth 1 \cite{sylenth1} and place them as the generator of ours, called \code{LoopyPreset}.

The notes and pattern cores are well illustrated by Fig. \ref{fig:pattern_core_notes}. On this piano roll, each small rectangle is a note, connected to a generator. For example, the one surrounded by the red rectangle is an eighth note. Our pattern core could be viewed as this piano roll that contains a bunch of notes.

The list of samples in Fig. \ref{fig:sample-cores} correspond to the list of sample cores in "LooPy". Each sample core is simply a recorded and processed waveform as shown in \ref{fig:sample-core}. Many drum kit elements (kick, snare, hi-hat, etc.) and special sound effects (stabs, sweeps, ambience, etc.) are commonly made as samples that could be easily dragged and dropped into a track.

\begin{figure}[!htbp]
    \includegraphics[width=8cm]{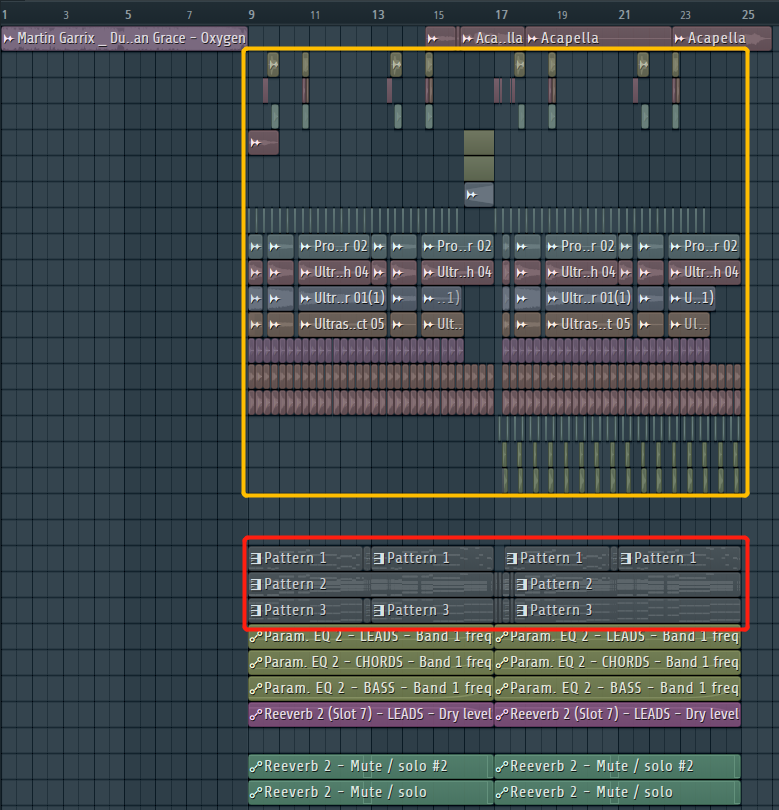}
    \caption{The playlists of a track}
    \label{fig:track_samples_patterns}
\end{figure}

\begin{figure}[!htbp]
    \includegraphics[width=8cm]{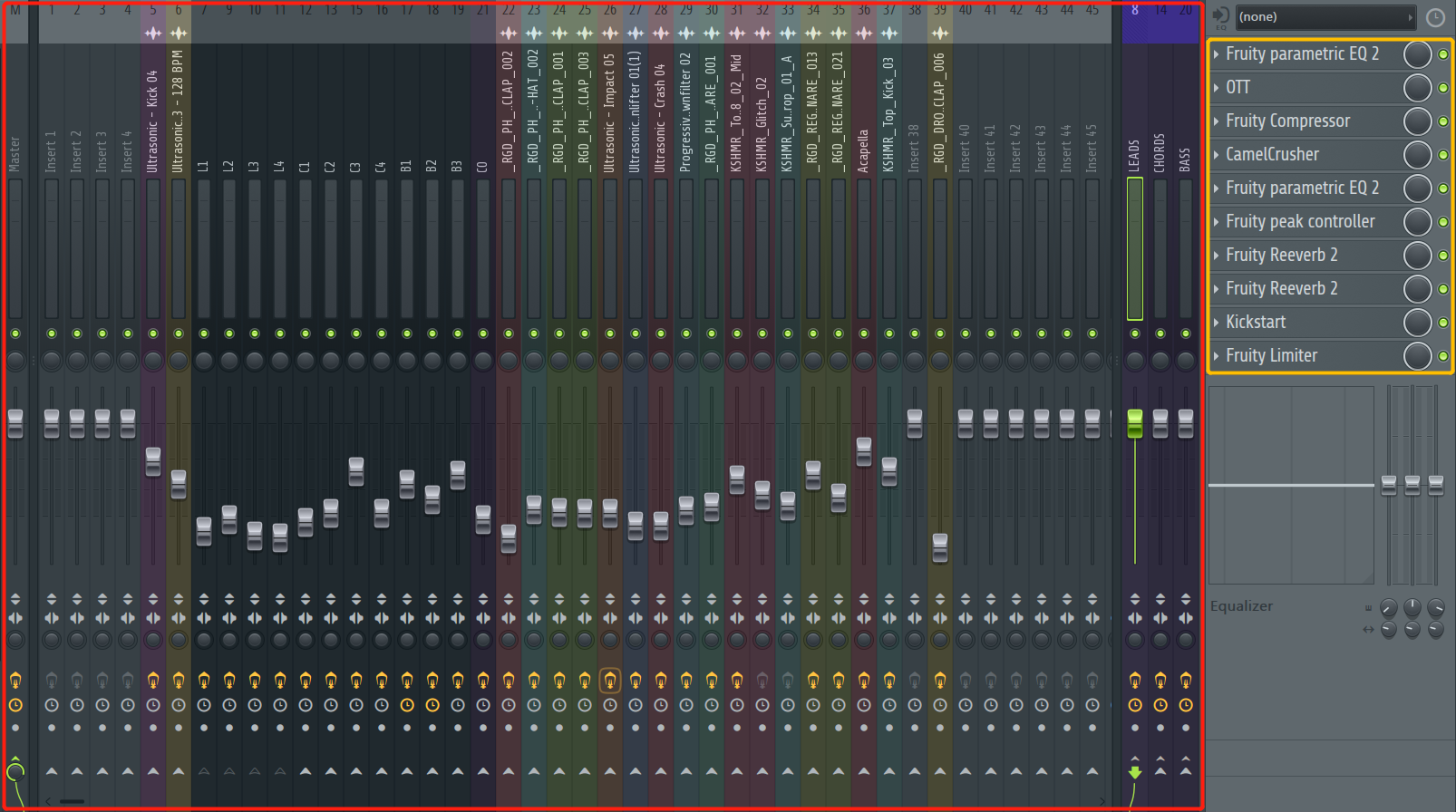}
    \caption{The mixer channels and effects of a track}
    \label{fig:channels_effects}
\end{figure}

In FL Studio, the patterns (surrounded by the red box) and samples (surrounded by the yellow box) are placed in the track as shown in Fig. \ref{fig:track_samples_patterns}. For "LooPy", when a \code{LoopyPatternCore} or \code{LoopySampleCore} instance is given the position in the track, such as "the beginning of the $8$-th bar", we would a \code{LoopyPattern} or \code{LoopySample} instance created.

As for rendering, each \code{LoopyChannel} corresponds to a mixer channel in Fig. \ref{fig:channels_effects} circled by the red box and each effect maps to an effect plugin circled by the yellow box. Note that one major difference between "LooPy" and FL Studio is that we route the audio of a pattern into a channel while the latter one routes the audio of a generator into a channel. Currently, we have no evidence on either ours or FL's mode is better, but our setup is subject to change according to user feedbacks in the future.

\section{Sidechaining}
\label{appendix:sidechain}

\begin{figure}[!bp]
 \centering
 \subfloat[][instrument waveform]{\includegraphics[height=3.12cm,keepaspectratio]{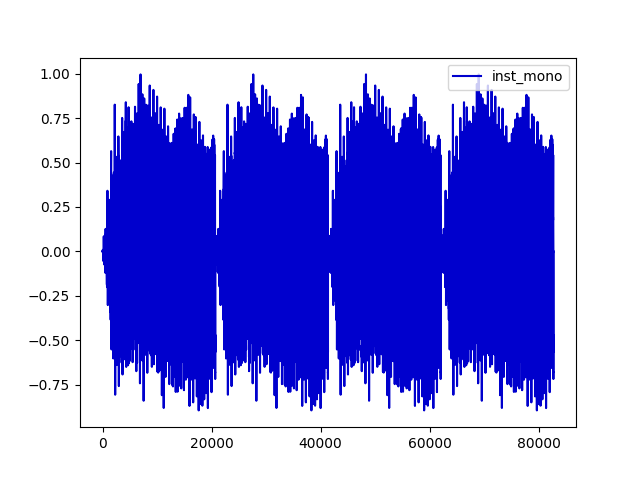}\label{figure1}}\hspace{0.02cm}
 \subfloat[][sidechain envelope]{\includegraphics[height=3.12cm,keepaspectratio]{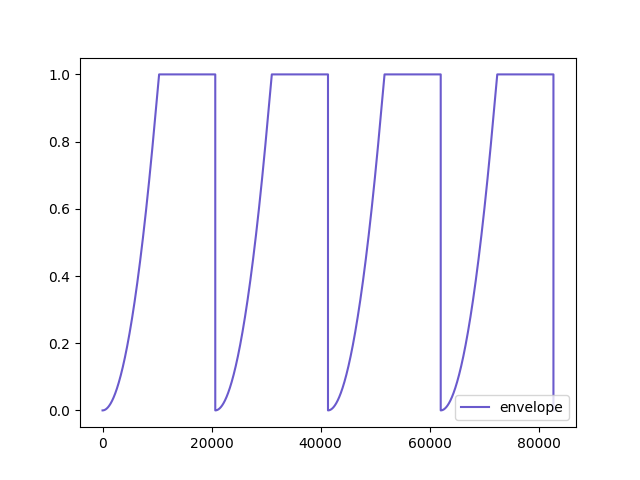}\label{figure2}}\null\\
 \hfill
 \subfloat[][kick drum waveform]{\includegraphics[height=3.12cm,keepaspectratio]{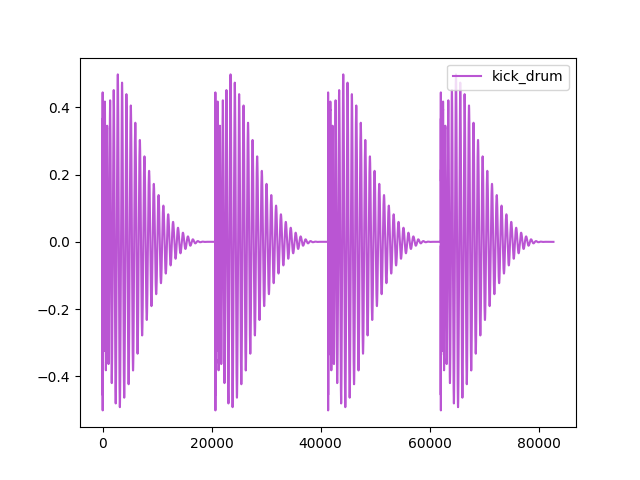}\label{figure3}}\hspace{0.02cm}
 \subfloat[][mix waveform]{\includegraphics[height=3.12cm,keepaspectratio]{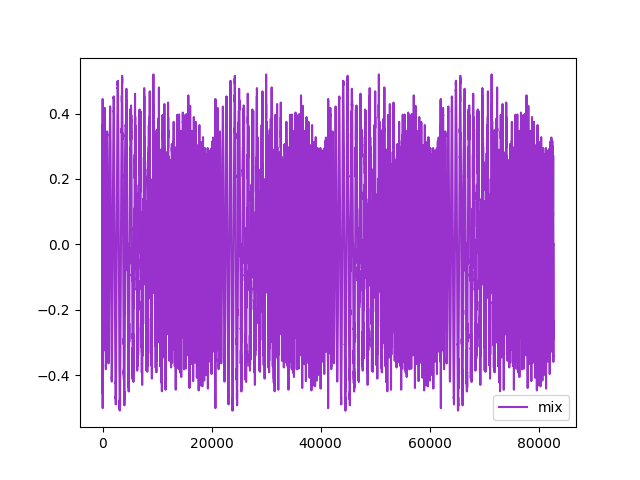}\label{figure4}}\null
 \caption{Illustration of sidechain.}
 \label{fig:sidechain}
\end{figure}

It is worth noting that sidechaining is an essential part in the mixing of EDM songs. The sidechain envelope is applied to avoid clashing between the kick drum and instruments, in order to maintain the punch in the kick and the cleanness of the entire mix. In Fig. \ref{fig:sidechain}, we give an example of the sidechain effect on a chord synth when mixing it with the kick drum. We only plot the mono waveform here for clarity\footnote[6]{Also, the $y$-axis of the four sub-figures are not perfectly consistent since we balanced the volume before mixing the waveforms.}. The next paragraph gives a rigorous definition of sidechaining.

Given 1 bar, sample rate equal to 44100 Hz, BPM equal to 128 and time signature equal to 4/4, we have 4 beats in the bar. The total number of samples in a beat is $$N_{beat}=int(\frac{60\times44100}{128})\approx20671$$, and that in a bar $$N_{bar}=int(\frac{4\times60\times44100}{128})\approx82687$$. First, we introduce the naive sidechain formula. Mathematically, denote the envelope by $e$, the instrument waveform by $y_{inst}$ and the kick drum waveform by $y_{kick}$. We can view $e$ as a periodic function from $\mathbb{N}$ to $[0, 1]$, which is non-decreasing in each of its period of length $N_{beat}$. In each period, the index where it reaches $1$ and the order of growth before it reaches $1$ are all parameters where users can tune according to the style they want to achieve, since the shape of the sidechain envelope directly affects the groove of the mix. As an array, $e$ has shape $(N_{bar},)$. Both the waveforms $y_{inst}$ and $y_{kick}$ have shape $(N_{bar},2)$ since they are stereo. we can write the mix waveform $y_{mix}$ as $$y_{mix}=y_{kick}+e\cdot y_{inst}$$, where $\cdot$ is the element-wise product\footnote[7]{In the implementation, the 2D (stereo) waveform array is multiplied by the envelope by broadcasting.}.

Precisely, we don't want everything $100\%$ sidechained since that would make the mix too swingy. We apply sidechain for all the instruments, each with a different mix level, to guarantee a professional groove. The mix level here is defined as $\alpha\in(0, 1)$ where $$y_{mix}=y_{kick}+\alpha\cdot e\cdot y_{inst}+(1-\alpha)y_{inst}$$. Empirically, $alpha\approx 0.8$ as advised by experts\footnote[8]{https://www.youtube.com/watch?v=awNv4-0uU5o}.

\end{document}